\documentclass[twoside,twocolumn,aps,pra,showpacs]{revtex4-1}
\usepackage[dvips]{color}
\usepackage[latin9]{inputenc}
\usepackage[english]{babel}
\usepackage{graphicx}
\usepackage[nice]{nicefrac}



\begin{document}
\selectlanguage{english}

\title{Strong superadditivity and monogamy of the R\'enyi measure of entanglement}

\author{Marcio F. Cornelio}

\email{mfc@ifi.unicamp.br}

\author{Marcos C. de Oliveira}

\email{marcos@ifi.unicamp.br}

\affiliation{Instituto de F\'isica Gleb Wataghin, Universidade Estadual de Campinas,
Caixa Postal 6165, CEP 13084-971, Campinas, S\~ao Paulo, Brazil}

\begin{abstract}
Employing the quantum R\'enyi $\alpha$-entropies as a measure of entanglement, we numerically find the violation of the strong superadditivity inequality for a system composed of four qubits and $\alpha>1$. This violation gets smaller as $\alpha\rightarrow 1$ and vanishes for $\alpha=1$ when the measure corresponds to the Entanglement of Formation (EoF). We show that the R\'enyi measure aways satisfies the standard monogamy of entanglement for $\alpha = 2$, and only violates a high order monogamy inequality, in the rare cases in which the strong superadditivity is also violated. The sates numerically found where the violation occurs have special symmetries where both inequalities are equivalent. We also show that every measure  satisfing monogamy for high dimensional systems also satisfies the strong superadditivity inequality. For the case of R\'enyi measure, we provide strong numerical evidences that these two properties are equivalent.

\end{abstract}
\pacs{03.67.Mn}
\maketitle

Quantum resources present several counterintuitive features allowing more efficient realization of classical and quantum communication tasks. Unfortunately it is hard to predict the way in which those features are distributed or extended. This is the case for several and important additivity problems, such as for the Holevo capacity of a quantum channel, minimal output entropy of a quantum channel, and the additivity of entanglement of formation (EoF) \cite{BVSW}, one of the most important entanglement measures.
These were shown by Shor \cite{shor} to be all equivalent to the
strong superadditivity (SS) \cite{VollWerner01} of the EoF. An entanglement measure $E$ satisfies SS if
\begin{equation}
E_{a_{1}a_{2}|b_1 b_2}\ge E_{a_{1}|b_1}+E_{a_{2}|b_2},
\label{eq:StrongSuperadditivity}
\end{equation}
meaning that, when Alice holds parties $a_{1}$ and $a_{2}$
and Bob holds parties $b_1$ and $b_2$, the entanglement
between Alice and Bob is larger than the entanglement between
$a_{1}$ and $b_1$ plus the one between $a_{2}$ and $b_2$. It is a very important relation since it is connected to the ability to extract arbitrary entangled states from
a standard one and to the ability to communicate classical information using a quantum channel. Moreover, if a measure $E$ is additive for pure states and extends for
mixed states through its convex roof, the SS implies the additivity of the measure
for mixed states as well. 
Althought the additivity of EoF was proved previously in some very particular cases \cite{VDC02},
recently Hastings  demonstrated in a remarkable work \cite{hastings} that once all of these
conjectures were equivalent they were also in general false due to
the existence of counterexamples for the minimal output entropy 
for sufficiently large dimensions of the Hilbert spaces involved.
Whether there is a violation of SS of EoF or not for lower dimensions is unknown.
Perhaps finding counterexamples for lower dimensions asks new Information Theoretical insights.

In this paper we derive an entanglement measure based on the $\alpha$-quantum R\'enyi entropy. For $\alpha>1$ we numerically find counter-examples violating the SS (\ref{eq:StrongSuperadditivity}) for four qubits systems, the smallest possible situation that SS can be written. This suggests that counter-examples for SS of EoF ($\alpha=1$) may exist for smaller dimensions. Moreover this measure also provides an important relation between SS and the so-called \textit{monogamy of entanglement} \cite{CKW}. The last is related to the way in that quantum correlation (entanglement) can be distributed between many parties. A measure of entanglement $E$ satisfying the monogamy relation with the Alice's subsystem $a$  and Bob's subsystems $b_1$ and $b_2$ must follow  
\begin{equation}
E_{a|b_{1}b_2}\ge E_{a|b_1}+E_{a|b_2}\label{eq:Emonogamy}.
\end{equation}
Important measures of entanglement, and particularly the EoF, fail to satisfy monogamy \cite{CKW}.
In some sense it seems that these two properties, in principle unrelated, strong
superadditivity and monogamy of entanglement, may actually be related and this
could be important for quantum information tasks since it would be also equivalent
to the other existent conjectures.
We start by discussing entanglement monogamy relations and we show how a second order monogamy relation implies the SS inequality independently of the measure of entanglement. Then we show that the R\'enyi measure, for $\alpha = 2$, satisfies the standard monogamy inequality for qubits. Numerically, we investigate the interrelation between these two inequalities using that measure. Interestingly, we find that violation of these two inequalities happens quite rarely but always simultaneously.
Thus we conjecture that SS violation of the R\'enyi measure for $\alpha = 2$
is necessary and sufficient for the second order monogamy violation.
After that we show how numerical methods can be used to find
violations of SS of the R\'enyi measure for $\alpha$ 
very close to one.

Monogamy of entanglement shows how
quantum correlation is special and different from classical
one - While classical correlation can be arbitrarily shared
with as many individuals as desired, quantum correlation cannot.
This impossibility of sharing quantum entanglement was first quantified
by Coffman, Kundu and Wooters (CKW) \cite{CKW}, through the squared concurrence $C^2$ as follows
\begin{equation}
C_{a|b_{1}b_{2}}^{2}(\rho_{ab_{1}b_{2}})\geq
C_{ab_{1}}^{2}(\rho_{ab_{1}b_2})+C_{a b_1 b_{2}}^{2}(\rho_{a b_1 b_{2}})
\label{eq:CKW-monogamy}
\end{equation}
for any pure or mixed state $\rho_{ab_{1}b_{2}}$ of a tripartite system built
of qubits, $a$, $b_{1}$ and $b_{2}$. 

Surprisingly, not all measures of entanglement
satisfy monogamy relations, for increased Hilbert space dimension and/or number of systems, in exception of the squashed entanglement \cite{KoashiWinter03}. 
Moreover  there exists a constraint in the CKW monogamy relation: it is true only when  $a$, $b_1$ and $b_2$ are qubits. In Ref. \cite{OV}, the authors extended its validity when $b_2$ is a $n$-level system, allowing them to prove the CKW monogamy for $N$-qubits, $C_{1|23...N}^{2}\geq C_{12}^{2}+C_{13}^{2}+...+C_{1N}^{2}$, as conjectured in Ref. \cite{CKW}. However, the inequality (\ref{eq:CKW-monogamy}) is not satisfied by increasing the dimension of $a$
\cite{Ou}.
In fact a measure of entanglement which is monogamous when the subsystem $a$ has higher
dimensions implies directly the SS as we now show.
Let us consider the case of subsystem $a$ being broken into two subsystems,
$a_{1}$ and $a_{2}$, and apply the monogamy relation (\ref{eq:Emonogamy}) 
again to obtain\begin{eqnarray}
E_{a_{1}a_{2}|b_{1}b_{2}} & \geq & E_{a_{1}|b_1}+E_{a_{2}|b_1}+E_{a_{1}|b_2}+E_{a_{2}|b_2}.\label{eq:2scMonogamy}\end{eqnarray}
We call this relation second order monogamy,
 whose meaning is similar to (\ref{eq:Emonogamy}): The amount of bi-partite entanglement
shared between $a_{1}\otimes a_{2}$ and $b_1\otimes b_2$ gives us an
upper bound to the sum of entanglement shared by $a_{1}$ and $b_1$,
$a_{2}$ and $b_1$, $a_{1}$ and $b_2$, and $a_{2}$ and $b_2$. This idea
can be generalized and we can obtain higher order monogamy relations
by successive applications of (\ref{eq:Emonogamy}). 
We are however more interested in
the fact that a measure $E$ satisfying this second order relation
(\ref{eq:2scMonogamy}) also satisfies the SS inequality (\ref{eq:StrongSuperadditivity}).
Remark however that by this reasoning it is not possible to show whether SS implies
the second order monogamy (\ref{eq:2scMonogamy}) directly or not. Instead, the SS is a necessary condition for satisfying monogamy for \textit{ any} measure of entanglement. Then we question if it is sufficient as well.

To investigate this point we choose the family of R\'enyi entropies which are
known to be additive. The quantum R\'enyi entropy of order $\alpha$ \cite{horodecki}
is defined as
\begin{equation}
R_{\alpha} = \frac{1}{1-\alpha} \log  \textrm{Tr} \rho^{\alpha},
\end{equation}
where $\alpha \geq  0$ and the logarithmic function will always be assumed to be
base 2 in this paper. In this way, for any pure bipartite system, 
the R\'enyi $\alpha$-entropy of one of the subsystems is a good and additive
measure of entanglement. The natural way to define the R\'enyi measure of entanglement, $\mathcal{R}_{\alpha}$, for a bipartite mixed state $\rho_{ab}$ is using
the convex roof reasoning of Ref. \cite{BVSW}. We consider the set $\mathcal{E}$ of all ensembles of pure states $|\varphi_{i}\rangle$ with weight $p_{i}$ realizing the state $\rho_{ab}$, $\rho_{ab}=\sum_{i}p_{i}|\varphi_{i}\rangle\langle\varphi_{i}|.$
For each ensemble, we can define an average value of $\mathcal{R}_{\alpha}$.
Then we define $\mathcal{R}_{\alpha}(\rho_{ab})$ as the minimal value of this
average over all the possible ensembles
\footnote{Due to the Schur-concavity of R\'enyi entropy, $\mathcal{R}_{\alpha}$
does not increase under deterministic LOCC \cite{ZB}.
However, the R\'enyi entropy, for $\alpha \le 2$ is known to be concave
only if the dimension of the space is
2 \cite{BR}. There is a counter-example 
to the concavity for dimension larger than 8 and $\alpha = 2$ \cite{BR}.
This implies that $\mathcal{R}_{\alpha}$ might increase on average under probabilistic
LOCC for higher dimensional systems \cite{vidal}.
},
\begin{equation}
\mathcal{R}_{\alpha}(\rho_{ab})=\min_{\mathcal{E}}\left\{ \sum_{i}p_{i}\mathcal{R}_{\alpha}(|\varphi_{i}\rangle)\right\} .
\end{equation}

In the case of two qubits, we can show an analytical expression for
$\mathcal{R}_{\alpha}$ for all $\alpha > 1$. For pure states, 
\begin{equation}
\mathcal{R}_{\alpha}(\rho_{ab})= \frac{1}{1-\alpha} \log \left[x^{\alpha} +(1-x)^{\alpha}\right],
\label{RforQubits}
\end{equation}
where $x=\nicefrac{(1+\sqrt{1-C^2})}{2}$ and $C$ is the concurrence \cite{BVSW}.
When $\alpha \rightarrow 1$ this formula goes to the usual one for the EoF \cite{wooters}.
To see that this relation is also true for mixed states, we must notice
that $\mathcal{R}_{\alpha}$ is a convex function of $C$ for $\alpha \geq 1$ and the ensemble
realizing the convex roof of concurrence is an ensemble composed of
states with the same value of $C$ \cite{wooters}. By
this construction, $\mathcal{R}_{\alpha}$ would be an additive measure if the SS was true.


Now we show that $\mathcal{R}_{2}$ satisfies the CKW monogamy for systems of $N$ qubits. 
Firstly we consider the case of a $N$-partite pure state. Noticing that Eq. (\ref{RforQubits})
simplifies for $\alpha=2$,
we can write the $\mathcal{R}_{2}$  between the
subsystem 1 and the other $(N-1)$ subsystems as $
\mathcal{R}^{1|23...N}_{2}=-\log\frac{(2-C_{1|234...N}^{2})}{2}\ge-\log\frac{(2-\sum_{i}C_{1i}^{2})}{2},$
 where the second inequality comes from the CKW monogamy \cite{CKW,OV}.
The entanglement between each two subsystems is given by eq. (\ref{RforQubits}).
Then, if we can show\begin{equation}
-\log\frac{(2-\sum_{i}C_{1i}^{2})}{2}\ge-\sum_{i}\log\frac{2-C_{1i}^{2}}{2},\label{eq:small-Inequality}\end{equation}
we obtain the CKW monogamy for $\mathcal{R}_2$.
But the inequality (\ref{eq:small-Inequality}) is equivalent to\begin{eqnarray}
0 & \le & \frac{1}{2^{2}}\sum_{i\neq j}\frac{1}{2!}C_{1i}^{2}C_{1j}^{2}-\frac{1}{2^{3}}\sum_{i\neq j\neq k}\frac{1}{3!}C_{1i}^{2}C_{1j}^{2}C_{1k}^{2}\nonumber \\
 &  & +\frac{1}{2^{4}}\sum_{i\neq j\neq k\neq m}\frac{1}{4!}C_{1i}^{2}C_{1j}^{2}C_{1k}^{2}C_{1m}^{2}\nonumber \\
 &  & -\frac{1}{2^{5}}\sum_{i\neq j\neq k\neq m\neq n}\frac{1}{5!}C_{1i}^{2}C_{1j}^{2}C_{1k}^{2}C_{1m}^{2}C_{1n}^{2}+...\label{eq:Big-inequality}\end{eqnarray}
It is easily seen that this inequality is always true since each negative term is always smaller than its preceding positive one. Since it implies the CKW monogamy, we have proved our claim. The result generalizes for mixed states by straightforward
use of the definition of $\mathcal{R}_2$ as a convex roof and the fact
the monogamy is true for pure states.


The R\'enyi measure of entanglement does
not satisfy the SS only in some very particular cases.
Numerically, we were able to find counter-examples for $\alpha$ very close to one
with systems of only four qubits (Fig. 1). The violation is smaller as $\alpha \rightarrow 1$
and vanishes for the case of EoF (Fig. 2). These counter-examples suggest that counter-examples
to the additivity of EoF and Holevo Capacity may exist for smaller dimensions.
Furthermore, in the particular case $\alpha = 2$,
we could not find any violation of monogamy inequality
not corresponding to the violation of the SS as well. 
In fact all states numerically found where this violation occurs are such that two of the bipartite entanglement appearing in right side of Eq. (\ref{eq:2scMonogamy}) vanish, being thus equivalent to the SS inequality (\ref{eq:StrongSuperadditivity}). 
Thus we conjecture that SS is necessary and sufficient for monogamy.

\begin{figure}
\includegraphics[scale=0.55]{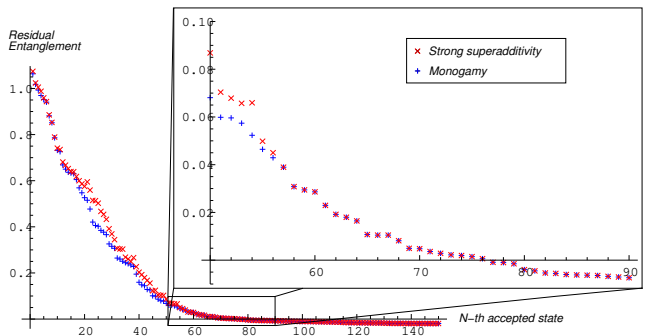}
\caption{Evolution of the minimization. Each point is a new state
found with a smaller residual entanglement than the previous one.
The total number of states generated in this example was about 10
thousands. One can check that the residues of monogamy and strong
additivity start to coincide when they are close to zero.
}
\label{Flo:graph}
\end{figure}

Violations of inequalities (\ref{eq:StrongSuperadditivity}) and (\ref{eq:2scMonogamy})
are not easy to find. For the case $\alpha = 2$, we were not able to find that
choosing 50 million pure states randomly (according to the Haar measure),
what takes about a week of computing time on a standard PC.
To find it we had to employ a simple Monte Carlo minimization algorithm. The function to be minimized is the difference between the first and the second members of (\ref{eq:StrongSuperadditivity}) and (\ref{eq:2scMonogamy}), also called residual entanglement \cite{CKW}. So there are two residual entanglements, one for SS (\ref{eq:StrongSuperadditivity}) and one for monogamy (\ref{eq:2scMonogamy}). The algorithm works as follows. First, we choose randomly a state as seed and we fix a distance $\delta$. Then we look randomly for a state with smaller residual entanglement within a distance (trace distance) $\delta$ from the seed. We also use a counter to count the number of random states generated until we find a state with smaller residual entanglement. Always when we find it we reset the counter and start the searching from this state as a new seed. When the counter gets some large value (one thousand is usually large enough) we divide by 2 the distance $\delta$ from the seed and reset the counter. When the distance gets smaller than $10^{-4}$ we stop (this is sufficient to get a precision of order $10^{-8}$). A standard PC can run this in some minutes for four qubits systems and the results are very reasonable. 
Fig. \ref{Flo:graph} shows the progress of the algorithm for one particular case.

With this method, the algorithm finds a vanish residual entanglement on 70\% of the runnings
and a negative residual entanglement of -0.0197 on the remaining 30\%. This state, which we call $|\psi_{vio}\rangle$, has a reduced density matrix $\rho_{a_{1}a_{2}}$ with eigenvalues $\left\{ 0.66,0.14,0.14,0.06\right\} $
and has a considerable entanglement, $\mathcal{R}_2=1.06$, between $a$ and $b$.
The density matrix $\rho_{a_{1}b_1}$ and $\rho_{a_{2}b_2}$ has
eigenvalues $\left\{ 0.997,0.003,0,0\right\} $, that is, they are
almost pure. So $|\psi_{vio}\rangle$ is very close to a product state
of the form $|\psi_{a_{1}b_1}\rangle\otimes|\psi_{a_{2}b_2}\rangle$.
The entanglement between the subsystems $a_{1}$ and $b_1$, and $a_{2}$
and $b_2$ are all equal to 0.54. The entanglements between all the other qubits
vanish. Therefore, these states can be characterized by showing entanglement only between the components relevant to the SS inequality and been close to product states of the subsystems 1 and 2. 


We also made an extensive numerical test to check if all states violating monogamy have these properties.
Using the search algorithm described, we obtain a sequence of states
forming a path from an initially random state to one of maximum
violation of (\ref{eq:2scMonogamy}). The states of this path start
to violate monogamy at the same point that they violate SS 
and the value of violation is always the same (see Fig. \ref{Flo:graph}), confirming that two bipartite entanglements of (\ref{eq:2scMonogamy})
vanish. Furthermore, during the process, thousands of random states
are generated near this path and tested. With this method we tested
more than 3 million states in many different runnings of the algorithm.
In order to check this more carefully, we made a modification in
the algorithm for not staying always near this path. When we get inside
the region of states having negative residual entanglement, we stop
to decrease the distance and start a random walk in that region. With
this modified method, we checked more 4 million states and all
of them have the same residual entanglement for monogamy and strong
superadditivity inequalities. Therefore, in the case of the
$\mathcal{R}_2$ for four qubit systems,
we conjecture that states violating the monogamy inequality (\ref{eq:2scMonogamy})
are the ones which also violate the SS (\ref{eq:StrongSuperadditivity}) as well. 

\begin{figure}
\includegraphics[scale=0.72]{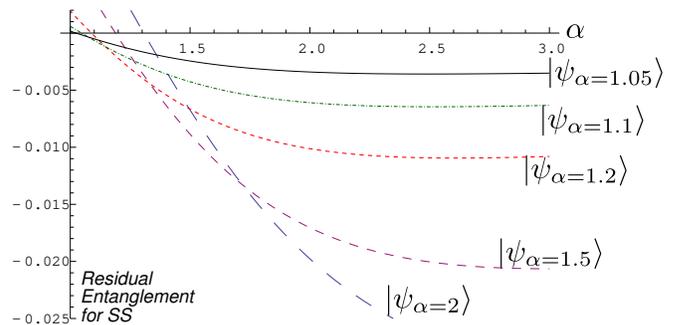}
\caption{Violation of SS for some states numerically found.
These states are found successively minimizing the residual entanglement for SS for
$\alpha = 2$, 1.5, 1.2, 1.1 and 1.05.}
\label{FigSS}
\end{figure}

Finding violation for the SS for $\alpha$ close to one is more
difficult. The violation gets very small and the best strategy is to use  a recurrence procedure. Instead of starting our searching  with a random state, we start it with the state
which maximally violates SS for $\alpha = 2$ as a seed, but run the algorithm for minimizing the residual entanglement of SS with $\alpha = 1.5$ starting with a smaller distance of $10^{-2}$ and leave it decreasing until $10^{-8}$. Then we go successively to $\alpha = 1.2$, 1.1, 1.05,...
and soon on. The progress of this process can be seen in Fig. \ref{FigSS} and illustrates
how the violation of SS vanishes as $\alpha \rightarrow 1$.
With this method, we found violation for $\alpha = 1.002$ of order of $10^{-6}$.
For large $\alpha$, the violation saturates to a value depending on the state.
These counter-examples strongly suggest that there are counter-examples to the SS of R\'enyi
measure for all $\alpha > 1$.

 We have made an extensive search for counter-examples to SS for $\alpha=1$ using these methods. As the numerical methods were efficient for finding counter-examples for almost every $\alpha$, we have a strong indication that there are no violation to SS of EoF for four qubits systems.
Despite that, the existence of counter-examples to SS for $\alpha$ close to one
at these very small dimensions suggest that there can be counter-examples to SS of EoF, and for all the others equivalent additivity questions,
for reasonable smaller dimensions then the ones necessary in the Hastings's counter-examples.
It is important to remember that his counter-examples were inspired by previous ones
of Hayden and Winter \cite{HW} for the minimal R\'enyi entropy output of a quantum channel.
So the counter-examples found here can be considered as a good indication of the
existence of analogous ones for the EoF.
The existence of such counter-examples,
for smaller dimensions,
would have great implications for quantum information.
It would imply that the superadditivy of the Holevo capacity and the subadditivity of
EoF can be used to improve the ability of communication over a quantum channel and
of the ability of forming states from a standard resources like EPR-pairs in
more practical
and simpler
situations.

In this work, we connected the properties of monogamy
and additivity of entanglement using the R\'enyi measure. We show
that this measure satisfies the standard monogamy inequality
for the particular case $\alpha = 2$. We also show that
the second order monogamy (\ref{eq:2scMonogamy}) implies the
strong superadditivity (\ref{eq:StrongSuperadditivity}).
Again in the case of $\alpha = 2$, we found numerically
that the inequalities (\ref{eq:StrongSuperadditivity}) and (\ref{eq:2scMonogamy})
are violated rarely, but always simultaneously and with the same magnitude.
Further, we provided strong numerical support for conjecturing
that the violation of monogamy inequality (\ref{eq:2scMonogamy}) is related
to the strong superadditivity (\ref{eq:StrongSuperadditivity}) violation for the
R\'enyi measure of order 2. This approach allowed to find more counter-examples
for SS as $\alpha$ gets closer to one. Also,
there are counter-examples to the SS of the R\'enyi measure for every $\alpha > 1$.
The violation of SS becomes very small as $\alpha \rightarrow 1$ and vanishes for
$\alpha = 1$.

The results here can help to understand why EoF turns out to be
non-additive.
The counter-examples found can stimulate the research of new counter-examples to
the additivity of EoF at small dimensions. 
Once the numerical methods employed are very simple, they can certainly be improved. This fact opens the possibility of numerical searching for such counter-examples for small dimensions, larger than 4 by 4.
Since additivity and monogamy seem to be connected through our findings we expect that it may shed some light in the understanding of the way in which entanglement is distributed.
The R\'enyi measure introduced here certainly take an important role in this questioning
as well as the question of
finding new counter-examples to the additivity of EoF.

We acknowledge T. R. de Oliveira for fruitful discussions. This
work was supported by FAPESP and CNPq through
the National Institute of Science and Technology of Quantum Information (INCT-IQ).

\end{document}